# Derivation of transformation equations for the parameters that characterize a plane acoustic wave without using phase invariance and Lorentz-Einstein transformation

Bernhard Rothenstein, "Politehnica" University of Timisoara, Physics Department, Timisoara, Romania

*Abstract. We show that the transformation equations for the parameters that characterize a plane acoustic wave: period, (frequency), wave vector, wave length and phase velocity can be derived without using phase invariance and Lorentz-Einstein transformation.*

1. **Introduction.**
Moller[1] derives transformation equations for the parameters introduced in order to characterize a plane acoustic wave. Resnick[2] and Rosser[3] present the same problem in the case of a plane electromagnetic wave propagating in a vacuum. The mentioned parameters are unit vector ***n*** that characterizes the direction in which the wave propagates with phase velocity ***w***, frequency ***ν*** and wave length ***λ***. The parameters are measured in the K(XOY) reference frame using rulers and synchronized clocks following the synchronization procedure proposed by Einstein[4]. The same parameters measured by observers from the K'(X'O'Y') reference frame are ***n'***, ***w'***, ***ν'*** and ***λ'***. The axes of the two reference frames are parallel to each other, K' moves with velocity ***V*** relative to K in the positive direction of the common OX(O'X') axes. At the origin of time in the two reference frames the origins O and O' of the two frames are located at the same point in space. Deriving transformation equations for the parameters mentioned above the authors invoke the invariance of the phase of a wave and perform the Lorentz-Einstein transformations of the space-time coordinates used in order to express the phase[1,2,3]. This approach leads to transformation equations for all the parameters.
The transformation equation for $\nu$ and $\nu'$ can be derived without using phase invariance special relativity being involved only via the time dilation effect. It holds exactly when the involved periods are very small.[5]
We show that the transformation equations for all the parameters can be derived without using phase invariance and Lorentz-Einstein transformations.

2. **Transformation equations for the parameters that characterize a plane acoustic wave.**
Detected from K, the parallel rays of the plane acoustic wave propagate along a direction that makes an angle $\alpha$ with the positive direction of the OX axis. One of the rays arrives at the origin O when clock $C_0(0,0)$ located there reads t=0. After a time *t* of propagation the ray arrives at the point *M(x,y)*. The phase of the harmonic oscillations taking place in the wave is $\Psi$ when detected from K and $\Psi'$ when detected from K'. The invariance of the phase requires that

$$\nu\left(t - \frac{x\cos\alpha + y\sin\alpha}{w}\right) = \nu'\left(t' - \frac{x'\cos\alpha' + y'\sin\alpha'}{w'}\right). \qquad (1)$$

The problem is to derive transformation equations for the parameters characterizing the wave, knowing that *x,y,t* and *x',y't'* transform in accordance with the Lorentz-Einstein transformations for the space-time coordinates of the same event. Even if the derivations are straightforward they are not very transparent as they are rather formal and obscuring the physics behind wave relativity.[6]

We consider that when we speak about a given physical quantity we should mention the observer who performs the measurement, where and when he performs it and the measuring devices he uses. What we measure in a Doppler Effect experiment is time intervals[5] (periods) which enable us to calculate the corresponding frequencies. The frequencies $\nu$ and $\nu'$ in (1) are instantaneous frequencies defined as the time derivatives of the phase

$$\nu = \frac{d\Psi}{dt}. \tag{2}$$

We present a simple and transparent derivation of the transformation equation for frequency (Doppler formula). Consider the ray that arrives at the origin O of K when clock $C(0,0)$ reads $t_e$. After a given time of propagation the ray arrives at the location of a clock $C(x = r\cos\alpha, y = r\sin\alpha)$ that reads $t_r$. We define the location of the clock using Cartesian $(x,y)$ and polar $(r,\alpha)$ space coordinates. It is obvious that the readings of the two clocks are related by

$$t_r = t_e + \frac{r}{w}. \tag{3}$$

We allow for a small change in the readings of the two clocks. The result is

$$dt_r = dt_e + \frac{dr}{w}. \tag{4}$$

Let $r'(x' = r'\cos\alpha', y' = r'\sin\alpha')$ be an observer of the K' reference frame and $C'(x' = r'\cos\alpha', y' = r'\sin\alpha')$ a clock commoving with him. By definition

$$\frac{dr}{dt_r} = V\cos\alpha \tag{5}$$

represents the radial component of the instantaneous velocity of *R'* and of his clock, relative to K. With (5) equation (4) leads to

$$\frac{dt_e}{dt_r} = 1 - \frac{V}{w}\cos\alpha. \tag{6}$$

Equation (6) holds for each value of $t_e$ and we are specifically interested in the case $t_e=0$. By definition $dt_e$ represents a proper time interval measured as a difference between the readings of the same clock $C_0(0,0)$. The clock $C'(x' = r'\cos\alpha', y' = r'\sin\alpha')$ records a change in its reading $dt'_r$ related to $dt_r$ by the time dilation formula

$$dt_r = \frac{dt'_r}{\sqrt{1 - \frac{V^2}{c^2}}} \tag{7}$$

because $dt'_r$ is a proper time interval, $dt_r$ is a non-proper time interval. With (7) formula (6) becomes

$$\frac{dt_e}{dt_r^{'}} = \frac{1 - \frac{V}{w}\cos\alpha}{\sqrt{1 - \frac{V^2}{c^2}}} \ . \tag{8}$$

We can consider that $dt_e$ represents a proper time interval during which an observer $R(x = r\cos\alpha, y = r\sin\alpha)$ of the K frame receives and emits successive wave crests of the wave, with $dt_r^{'}$ representing a proper time interval during which an observer $R'(x' = r'\cos\alpha', y' = r'\sin\alpha')$ of K' receives the same wave crests of the same wave. Knowing that in a Doppler Effect experiment we compare the period at which two successive wave crests are emitted, measured in the rest frame of the source (K), with the period at which they are received measured in the observer's rest frame (K'), we can consider that (8) describes the Doppler Effect.[7] Expressed as a function of frequencies ($\nu = \frac{1}{dt_e}$!!, $\nu' = \frac{1}{dt_r^{'}}$!!) equation (8) becomes

$$\nu' = \nu \frac{1 - \frac{V}{w}\cos\alpha}{\sqrt{1 - \frac{V^2}{c^2}}} \tag{9}$$

recovering the result obtained via the phase invariance and Lorentz-Einstein transformation approach.[1] Our derivation of (9) reveals the following physical facts behind it:
- It holds exactly only in the case of "very small periods" ("very high frequencies"), small enough that it can be considered that during the time intervals $dt_e$ ($dt_r^{'}$) the angle $\alpha$ does not change, with $R'$ receiving two successive wave crests being located at the same point in space. Relativists say that we make the "locality assumption" in the period measurement by a moving observer.[8] We find a similar situation in the case of an accelerating observer whose velocity changes during the reception of two successive wave crests.[9]
-The Lorentz-Einstein transformation performed in Moller's approach[1] relates the space-time coordinates of the event "observer R' detects two successive wave crests from the same point in space".
-Equation (9) holds only for the ray which arrives at the origin O of the K frame when clock $C_0(0,0)$ reads a zero time.

Considering the same Doppler Effect experiment from the K' reference frame, where R' is in a state of rest we obtain

$$\nu = \nu' \frac{1 + \frac{V}{w'}\cos\alpha'}{\sqrt{1 - \frac{V^2}{c^2}}} \tag{10}$$

Combining (9) and (10) we obtain

$$\frac{\cos\alpha'}{w'} = \frac{\cos\alpha - \frac{V}{c^2}w}{w(1 - \frac{V}{w}\cos\alpha)}. \tag{11}$$

Combined with (10) equation (11) leads to

$$\frac{v'\cos\alpha'}{w'} = \frac{v}{w}\frac{\cos\alpha - \frac{V}{c^2}w}{\sqrt{1 - \frac{V^2}{c^2}}}. \tag{12}$$

Combining (9) and (10) we obtain

$$\frac{v'}{w'}\sin\alpha' = \frac{v}{w}\sin\alpha. \tag{13}$$

Introducing the concept of propagation vector

$$\mathbf{k} = \frac{v}{w}\mathbf{n} \tag{14}$$

in K and

$$\mathbf{k}' = \frac{v'}{w'}\mathbf{n}' \tag{15}$$

in K', the equations derived above become

$$k'_x = k\frac{\cos\alpha - \frac{V}{c^2}w}{\sqrt{1 - \frac{V^2}{c^2}}} = \frac{k_x - \frac{V}{c^2}v}{\sqrt{1 - \frac{V^2}{c^2}}} \tag{16}$$

$$k'_y = k_y. \tag{17}$$

The angles $\alpha$ and $\alpha'$ that characterize the direction of propagation transform as

$$tg\alpha' = \frac{k'_y}{k'_x} = \frac{\sqrt{1 - \frac{V^2}{c^2}}\sin\alpha'}{\cos\alpha' - \frac{V}{c^2}w} \tag{18}$$

with the phase velocity transforming as

$$w' = \frac{v'}{k'}\frac{w - V\cos\alpha}{\sqrt{(\cos\alpha - \frac{V}{c^2}w)^2 + (1 - \frac{V^2}{c^2})\sin^2\alpha}}. \tag{19}$$

Finally, we obtain that the wave length ($\lambda = \frac{w}{v}, \lambda' = \frac{w'}{v'}$) transforms as

$$\lambda' = \frac{w'}{v'} = \lambda\frac{\sqrt{1 - \frac{V^2}{c^2}}}{\sqrt{(\cos\alpha - \frac{V}{c^2}w)^2 + (1 - \frac{V^2}{c^2})\sin^2\alpha}}. \tag{20}$$

The important conclusion is that we have derived the transformation equations for the parameters that characterize a plane acoustic wave without using the phase invariance and the Lorentz-Einstein transformations. Special relativity is involved in our derivations via the time dilation formula which is itself a direct consequence of the principle of relativity.

Equations (16), (17) and (9) presented as

$$v' = \frac{v - Vk_x}{\sqrt{1 - \frac{V^2}{c^2}}} \quad (21)$$

tell us that $(\frac{v}{c}, k_x, k_y)$ and $(\frac{v'}{c}, k'_x, k'_y)$ are, in our two dimensional approach, the components of a "wave three-vector" satisfying the relativistic identity

$$v't' - k'_x x' - k'_y y' = vt - k_x x - k_y y. \quad (22)$$

We recognise in the left hand side and in the right hand side of (22) the dot product of the "wave three-vector" defined above and the position vector in space-time *x',y,ct'* (*x,y,ct*). As we know the dot product of two "three vectors" is a relativistic invariant in accordance with the invariance of the phase. Corresponding components of two "three vectors" transform via the same transformation factors and so we should have

$$x' = \frac{x - Vt}{\sqrt{1 - \frac{V^2}{c^2}}} \quad (23)$$

$$y' = y \quad (24)$$

$$t' = \frac{t - \frac{V}{c^2}x}{\sqrt{1 - \frac{V^2}{c^2}}} \quad (25)$$

recovering the well known Lorentz-Einstein transformations.

**3.       Transformation equations for the parameters characterizing a plane electromagnetic wave**

An electromagnetic wave propagates in vacuum with the invariant phase velocity *c*. We obtain the corresponding transformation equations from those derived above by simply making *u=u'=c*, i.e.

$$tg\alpha' = \frac{\sqrt{1 - \frac{V^2}{c^2}} \sin\alpha}{\cos\alpha - \frac{V}{c}} \quad (26)$$

$$v' = v\frac{1 - \frac{V}{c}\cos\alpha}{\sqrt{1 - \frac{V^2}{c^2}}} = \frac{v - V\frac{v}{c}\cos\alpha}{\sqrt{1 - \frac{V^2}{c^2}}}. \quad (27)$$

Relation (27) suggests defining a vector

$$\mathbf{k} = \frac{\nu}{c}\mathbf{n} \tag{28}$$

the components of which transform as

$$k'_x = \frac{k_x - \frac{V}{c^2}\nu}{\sqrt{1-\frac{V^2}{c^2}}} \tag{29}$$

$$k'_y = k_y. \tag{30}$$

Expressed as a function of the projection of **k,** (27) becomes

$$\nu' = \frac{\nu - Vk_x}{\sqrt{1-\frac{V^2}{c^2}}} \tag{31}$$

Recovering the results obtained following the phase invariance strategy.[2,3]

**4.     Conclusions**

    We have shown that the transformation equations for the parameters that characterize a plane acoustic or electromagnetic wave can be derived without invoking the invariance of the phase. The transformation equation for the frequency holds exactly only in the case of very small period assumption (locality in the period measurement by a moving observer) and of course in the case of the plane wave assumption (very big source-receiver distance).

**5.     References**

[1]C.Moller, *The Theory of Relativity,* (Clarendon Press, Oxford, 1972) Ch.2.9

[2] Robert Resnick, *Introduction to Special Relativity,* (John Wiley and Sons, Inc. New York, London, Sydney, 1968) pp. 84-91

[3]W.G.V. Rosser, *An Introduction to the Theory of Relativity,* (Butterworth, London, 1964) pp. 158-160

[4]Hans C. Ohanian, *Special Relativity: A Modern Introduction,* (Physics Curriculum and Instruction, Inc.Lakeville USA, 2001) pp. 42-49

[5]Daniel R. Frankl, "General treatment of the Doppler Effect in special relativity," Am.J.Phys. **52**, 370-375 (1984)     `

[6]Asher Peres, "Relativistic telemetry," Am.J.Phys. **55,** 516-519 (1987)

[7]Kurtiss J. Gordon, "The Doppler Effect: A consideration of quasar red shifts," Am.J. Phys. **48**, 514-517 (1980)

[8]William Moreau, "Nonlocality in the frequency measurements of uniformly accelerating observers, Am.J.Phys. **60,** 561-564 (1992)

[9]Bernhard Rothenstein and Aldo De Sabata, "Frequency shifts for accelerating sources and observers," Eur.J.Phys. **19**, 569-574 (1998)